\documentclass[fleqn,10pt]{wlscirep}
\usepackage[utf8]{inputenc}
\usepackage[T1]{fontenc}
\usepackage{lineno}
\usepackage{upgreek}
\usepackage{xspace}

\title{Towards real-time oxygen sensing: From nanomaterials to plasma}

\author[1,2,3,$\ddag$]{Vinitha Johny}
\author[1,2,3,$\ddag$]{KV Chinmaya}
\author[4]{Muhammed Nihal CV}
\author[1]{Varghese Kurian}
\author[3,4,5]{G Mohan Rao}
\author[3,4,5,6]{Moumita Ghosh}
\author[1,2,3,7,8,9,$\ast$]{Siddharth Ghosh}
\affil[1]{Department of Science, Open Academic Research Council, Kolkata, India}
\affil[2]{Department of Science, Open Academic Research UK CIC, Cambridge, UK}
\affil[3]{International Center for Nanodevices, Centre for Nano Science and Engineering, Indian Institute of Science, Bengaluru, India}
\affil[4]{Department of Technology, Open Academic Research Council, Kolkata, India }
\affil[5]{Department of Technology, Open Academic Research UK CIC, Cambridge, UK}
\affil[6]{Advanced Technology Group of FEI Electron Optics,  Thermo Fischer Scientific, Eindhoven, Netherlands}
\affil[7]{Department of Applied Mathematics \& Theoretical Physics, University of Cambridge, Cambridge, UK }
\affil[8]{Maxwell Centre, Cavendish Laboratory, University of Cambridge, Cambridge, UK }
\affil[9]{St John's College, University of Cambridge, Cambridge, UK }
\affil[$\ddag$]{Equally contributing author}
\affil[$\ast$]{corresponding authors:  siddharth@openacademicresearch.org}

\begin{abstract}
    A significantly large scope is available for the scientific and engineering developments of high-throughput ultra-high sensitive oxygen sensors. 
We give a perspective of oxygen sensing for two physical states of matters -- solid-state nanomaterials and plasma. 
From single-molecule experiments to material selection, we reviewed various aspects of sensing, such as capacitance, photophysics, electron mobility, response time, and a yearly progress. 
Towards miniaturisation, we have highlighted the benefit of lab-on-chip-based devices and showed exemplary measurements of fast real-time oxygen sensing. 
From the physical-chemistry perspective, plasma holds a strong potential in the application of oxygen sensing. 
We investigated the current state-of-the-art of electron density, temperature, and design issues of plasma systems.
We also show a numerical aspects of low-cost approach towards developing plasma-based oxygen sensor from household candle flame. 
In this perspective, we give an opinion about a diverse range of scientific insight together, identifies the short comings, and opens the path for new physical-chemistry device developments of oxygen sensor along with providing a guideline for innovators in oxygen sensing. 
\end{abstract}

\begin{document}

    \flushbottom
    \maketitle

    
    \section{Introduction}

\par
\begin{figure*}[htp!]
        \centering
        \includegraphics[width=1.0\textwidth]{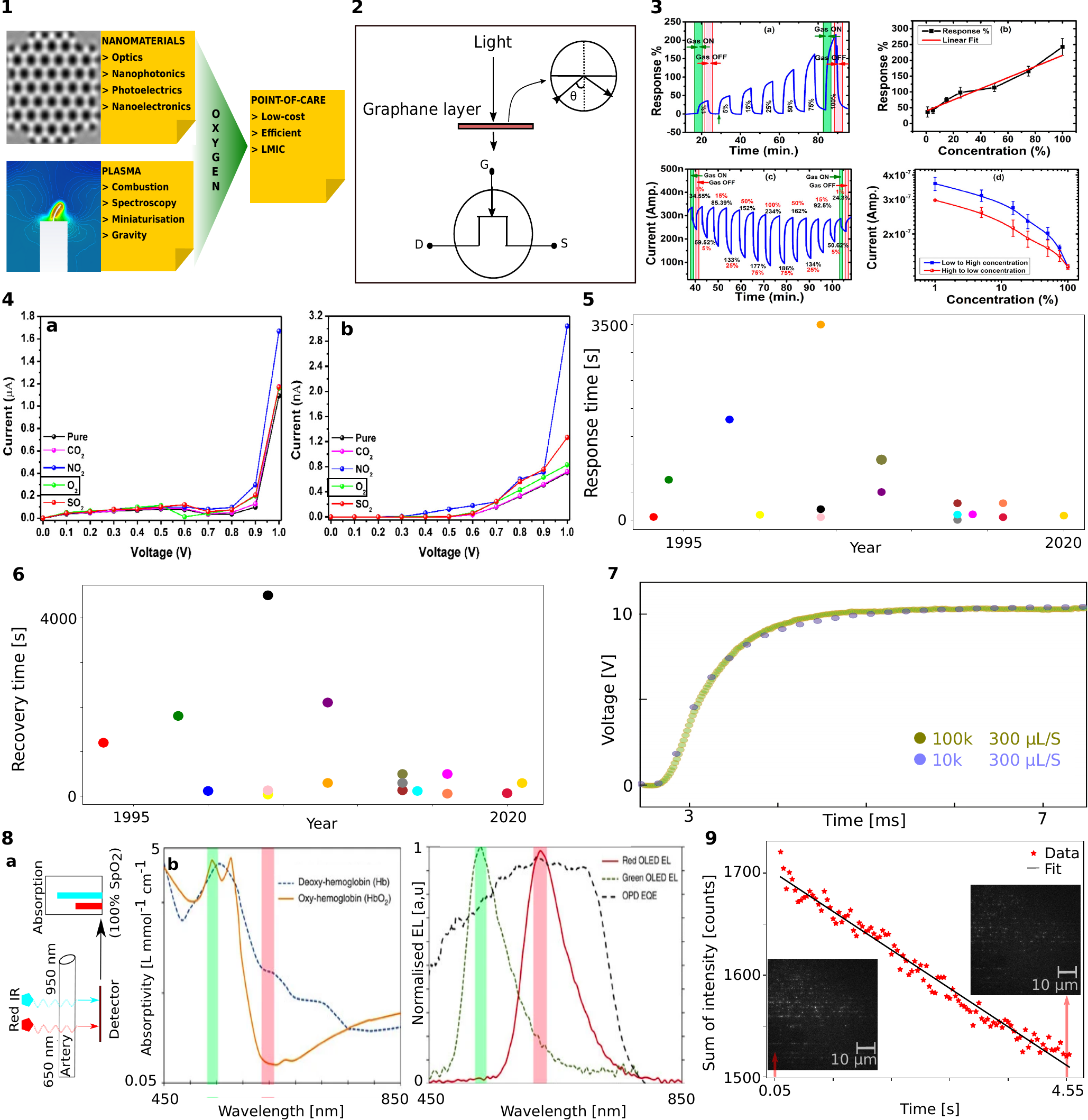}
        \caption{\textbf{1}, Overview of oxygen sensing with nanomaterials and plasma towards PoC. 
        \textbf{2}, Principle of graphene-based refractive index sensor.
        \textbf{3a} and \textbf{3b}, Dynamic response of sensor for oxygen and its linear ﬁt.
        \textbf{3c} and \textbf{3d}, Temporal current output with respect to concentration and its hysteresis (adapted from \cite{sakhuja20201t}). \textbf{4a} and \textbf{4b}, Current voltage response of SiBi for oxygen in contrast to other gases along one axis and  another axis, respectively (adapted from \cite{kumar2021first}). 
       \textbf{5,} Comparison of real-time voltage curve of GRIS at a collection frequencies of 10 kHz and 100 kHz (adapted from \cite{xing2012sensitive}).
        \textbf{6}, Evolution of response time of materials for oxygen sensor over the period of 1990 and 2020. 
        \textbf{7},  Variation of recovery time of nanomaterial-based oxygen sensor with respect to year of published papers. 
        \textbf{8a}, Principle of pulse oximetry sensor with two wavelengths of light. 
        \textbf{8b}, Absorptivity of oxygenated (orange solid line)  and deoxygenated (blue dashed line) haemoglobin in arterial blood as a function of wavelength (adapted from Figure 1 of \cite{lochner2014all}).
        \textbf{9}, Photooxidation of single fluorphores inside nanofluidic device (scale bar = 10 $\mu$m)}
        \label{Figure:1}
    \end{figure*}

 On-site oxygen sensing at the point of care (PoC) is an imperative and timely issue but is a non-trivial problem to solve from the technological perspective in spite of the well-established fundamental science. 
The limit and standard of detecting specific elements or molecules with ultra-high sensitivity among various PoC technologies is a continuous debate but it needs to be resolved immediately to address the current demand\cite{suleman2021point}.
State-of-the-art engineering techniques can be developed efficiently if we explore the hidden possibilities within every length scale and the phases of materials. 
In this perspective, we communicate our opinion about diverse methods of nanomaterials and plasma for highly sensitive and cost-effective oxygen sensors with an aim towards miniaturisation and PoC (Figure \ref{Figure:1}.1). 
Severe acute respiratory syndrome (due to COVID-19 or resistant pneumonia or air pollution) is life-threatening and a huge crisis in the low and middle-income counties where a few days of oxygen therapy cannot be supplied while diagnostics are undergoing \cite{talwar2015surgical,fiji2020development,howie2020development}). 
The motivation for this perspective includes the unavailability of concentrated oxygen and the unreliability of resultant data regarding the concentration of oxygen level in the blood for medical requirements, which is a continuous problem in other global south countries as well.
We could not find a low-cost oxygen sensor after performing a significant search in the commercial space for an effective low-cost oxygen sensor \cite{ahn2004disposable, li2010potential, baryeh2017introduction}.
Measuring oxygen levels in the blood and in the atmosphere should be abundantly available to developers and the public, like clinical thermometers as domestic diagnostic tools and PoC.
We investigate the compatibility of existing nanomaterial-based oxygen-sensing mechanisms in PoC applications.
It also emphasises how the sensitivity and response time of nanomaterial-based sensors become crucial in oxygen sensing in the light of optics and electronics.
Beyond the solid-state phase of matter, the electrically highly conductive state that is plasma will be indispensable in detecting molecular parameters due to the strong long-range electric and magnetic-field interactions. 
Hence, the latter part of the article discusses the physics of plasma in oxygen sensing.
We also demonstrate the first low-cost approach towards developing single-molecule nanofluidics devices and plasma-based oxygen sensors from a candle flame towards PoC realisations. 

\section{Oxygen sensing with nanomaterials}

\subsection{Nanoscale physical chemistry}
\par
Nanomaterials are famous for various shapes, high surface to volume ratios, and nontrivial physics,  \cite{cao2004nanostructures,nie2007nanotechnology,park2009silicon,rao2009graphene,sozer2009nanotechnology}, which have been a central attraction since 1959 \cite{feynman2018there}.
Basic principle of oxidation can be applied to engineer low-cost oxygen sensors, such as exposing lead to air creates lead-oxide; O$_2$ + 2Pb = 2PbO.
Due to the slow process of reaction time, this kind of materials are not suitable for immediate response.
We need fast response as well as fast recovery to design high-throughput oxygen sensors for PoC applications.

\subsection{Selection of nanomaterials} 

The main influencing parameters for oxygen sensing using nanomaterials are operating temperatures, structure of material under consideration, and the material being used for better recovery and response time.
\cite{sakhuja20201t} showed room temperature oxygen detection using liquid exfoliated TiS$_2$ nanosheet.
Figure \ref{Figure:1}.3, shows the TiS$_2$ nanosheet sensing performance toward oxygen. 
The response time and recovery time of oxygen sensing with the TiS$_2$ material is 78 s and 70 s respectively at room temperature.  
The associated hysteresis error from Figure \ref{Figure:1}.3d was found to be $\pm$3.42\%. 
TiS$_2$ nanomaterial holds better response and recovery time compared to previously published articles\cite{ZnOnanorod, ZnOoxygen}. 
For instance, ZnO-based sensors require 4500 s to recover, which is less than (almost half of) the reported values of \cite{sakhuja20201t}, and it cannot operate at the entire span of oxygen concentration. 
Kumar et al. demonstrated the use of SiBi-nanosheets as sensors for oxygen-containing gases\cite{kumar2021first}. 
They observed that the adsorption energy of oxygen molecules is -$1.10$ eV with work-function $4.22$ eV. 
This fact provides an indication of getting a relatively long recovery time for detaching the oxygen molecule from the SiBi surface.
With SiBi-nanosheets, they observed a fast recovery time of 297 s for oxygen.
Hence, these two are fast responding materials for oxygen gas-sensing and a viable solution for PoC oxygen-sensors.

Neri et al. demonstrated the use of a highly sensitive oxygen sensor based on Pt-doped In$_2$O$_3$ nanopowders at 200 $^\circ$C, which showed 60\% of sensitivity\cite{neri2005highly}.
Li et al. used ZnO nanowire transistors for oxygen sensing\cite{li2004oxygen}.
These sensors show a fast sensitivity with recovery time of 47 s at 200 ppm ethanol exposure. 
Monolayer-graphene, graphene transistors, TiO$_2$, and zeolite also showed potential to be effective oxygen sensors based on their sensitivity\cite{chen2011oxygen, bai2014titanium,pan2017application}.
Nanomaterials like TiO$_2$, with different sample preparation techniques and doped with different materials gave multiple options of oxygen sensing with respect to their response and recovery time. 
Bai et al. briefly talked about TiO$_2$ nanomaterials-based oxygen sensors\cite{bai2014titanium}. 
It opens up multiple dimensions towards PoC usage of single materials with different doped nanostructures at different temperatures with different fabrication methods.

Although oxygen sensitivity decreases with an increase in recovery time, the response time varies greatly for different nanostructures and their operating condition.
In Figure \ref{Figure:1}.5 and Figure \ref{Figure:1}.6, we show the history of sensitivity of nanomaterials in oxygen sensing.
As of now, the highest response and recovery time of nanomaterials-based oxygen sensing was found in 2020.
These advanced nanomaterials with improved sensitivity for oxygen sensing in the air and blood should enable the industry to scale up \cite{vanderkooi1986new,fidelus2007zirconia,chong2020mechanistic,wu2009ratiometric,pumera2010graphene,ghosal2018biomedical,katayama2020development,huynh2021nanomaterials}.
Oxygen can also be detected using several bulk instruments, such as AFM and x-rays to obtain fast results in a limited time. 
Researchers have used graphane, self-chargeable nano-bio-supercapacitors with microelectronic circuits to sense oxygen and blood plasma \cite{kim2020graphene,lee2021nano}. 
However, these techniques are limited to rapid PoC applications. 
One can think of reducing the bulk instruments and using the idea to create oxygen sensors for PoC devices.

\subsection{Physics of sensing towards point-of-care}

Single-molecule detection with optics is another avenue towards PoC devices. 
Fluorescence correlation spectroscopy (FCS) is suited for the quantitative determination of molecular oxygen within biological gases and fluids\cite{opitz2003single}.
FCS can't be used as PoC devices unless the bulkiness of the optical setup is miniaturised.  
Optical oxygen sensing can be designed based on the principle of fluorescence bleaching, blinking or quenching by oxygen\cite{mcdonagh2008optical}. 
Optical oxygen sensors depend on the use of a light source, a light detector, and luminescent material that reacts to light.
Ground state oxygen exists in the triplet state and helps in the easy transition of associated molecules to attain its triplet state, which makes it an efficient quencher.
The unpaired spins of oxygen can induce the excited state of the fluorescent molecule to undergo intersystem crossing from the singlet state to the triplet state\cite{kawaoka1967role, bergman1968rapid}.

Currently, optical sensors are heavily exploited in medical diagnostics to measure the saturation level of oxygen in the blood.
Xing et al. demonstrated a graphene-based refractive index sensor (GRIS) (principle of GRIS shown in Figure \ref{Figure:1}.2) can quickly and sensitively monitor changes in the local refractive index with a quick response time\cite{xing2012sensitive}.
It can be combined with microfluidic techniques, are an ideal material for fabricating biosensor devices, which are in high demand. 
The voltage vs time response of the GRIS sensor is shown in Figure \ref{Figure:1}.7.
Further research should focus on utilising these nanomaterials and optics by integrating them with nanophotonics, nanofluidic, and microfluidic devices that imprint the major tool for faster diagnostics and detection of single molecules, viruses, and pathogens as a biosensor\cite{mcrae2015programmable,yamanaka2016printable,zhang2019nanomaterials}.
In Figure \ref{Figure:1}.9, we show the flow of single fluorophores of Atto-488 excited with 488 nm CW laser inside 30 to 100 nm solid-state nanofluidic channels bleached over time due to photooxidation.
The decay rate of the intensity counts of fluorophores inside the nanofluidics channel is -18.83x.
Within 4.55 s we could acquire significant statistics of this information. 
We need such nanofluidic devices for PoC for faster monitoring of the single molecules\cite{ghosh2020single}.
Lab on chip for these devices is the key research in upcoming years, such as designing microfluidic oxygen sensor devices to study the long-term control and monitoring of chronic and cyclic hypoxia\cite{grist2015designing}.

The ongoing pandemic has generated an unprecedented demand for PoC devices for pulse oximetry.
The basic principle of pulse oximetry has been developing since 1876, and it is one of the imperative life saving innovations till date\cite{von1876quantitative,nicolai1930reizstromerzeugung,tremper1989pulse,miyasaka2021tribute}. 
From the war zone to the intensive care unit, pulse-oximetry plays a vital role, and the situation remained unchanged during the COVID-19 pandemic to measure the oxygen saturation level in human blood\cite{greenhalgh2021remote}.
Pulse-oximetry works on the principle of spectrophotometry by measuring oxygen saturation using two light sources -- typically 660 nm (red) and 940 nm (infrared) shown in Figure \ref{Figure:1}.8a.
The finding is 145 years old but it has not yet integrated with 13-year-old smartphone technologies, which uses 2 nm CMOS technology while pulse-oximetry still remained primarily bulky and un-calibrated.
Lochner et al.\cite{lochner2014all} used organic materials for pulse-oximeter instead of conventional expensive opto-electronic components\cite{jubran1999pulse}.
They used green (532 nm) and red (626 nm) organic light-emitting diodes (OLEDs) with an organic photodiode (OPD) sensitive at the aforementioned wavelengths.
These organic optoelectronic pulse oximetry sensors showed oxygenation with 2\% error.
The electroluminescence of green OLED and red OLED has a Lorentzian response with respect to wavelength and peaks at 500 nm and 640 nm, respectively.  
The quantum efficiency of the OPD is over 50\% for visible wavelengths and over 98\% for $\approx$640 nm shown in Figure \ref{Figure:1}.8b. 

These optoelectronics are inexpensive and have good flexibility that will allow researchers to modify the medical sensors in new shapes and sizes for point of care applications. 
Recently, Pipen et al. showed the relationship of SpO$_2$ and heart rates using commercial oximeters and smartwatches in patients with lung diseases\cite{pipek2021comparison}.
They found a correlation between the results of commercial oximeters and smartwatch devices while evaluating SpO$_2$ and heart rate measurements. 
Both devices have negligible statistical differences in the evaluation irrespective of skin colour, waist circumference, presence of wrist hair, and enamel nail.
There will be certain limitations of false readouts, and patients with hypoxemia who have high arterial oxygen tension levels may not be detected\cite{ralston1991potential,buckley1994pulse,jubran1999pulse,greenhalgh2021remote}. 
Today, we see many new advances in optical oxygen sensors which are entering to the market\cite{michelucci2019optical,browne2021accuracy,mocco2021pulse}.
However, detailed analyses, calibrations, and modelling should be continued to review the current findings, which will accelerate the dissemination.
The different concepts of oxygen sensing can be regarded as a catalyst for developing highly sophisticated oxygen sensing techniques in the relatively near future.

\section{Oxygen sensing with Plasma}
\subsection{Physical chemistry of plasma} 
\begin{figure}[htp!]
    \centering
        \includegraphics[width=0.8\textwidth]{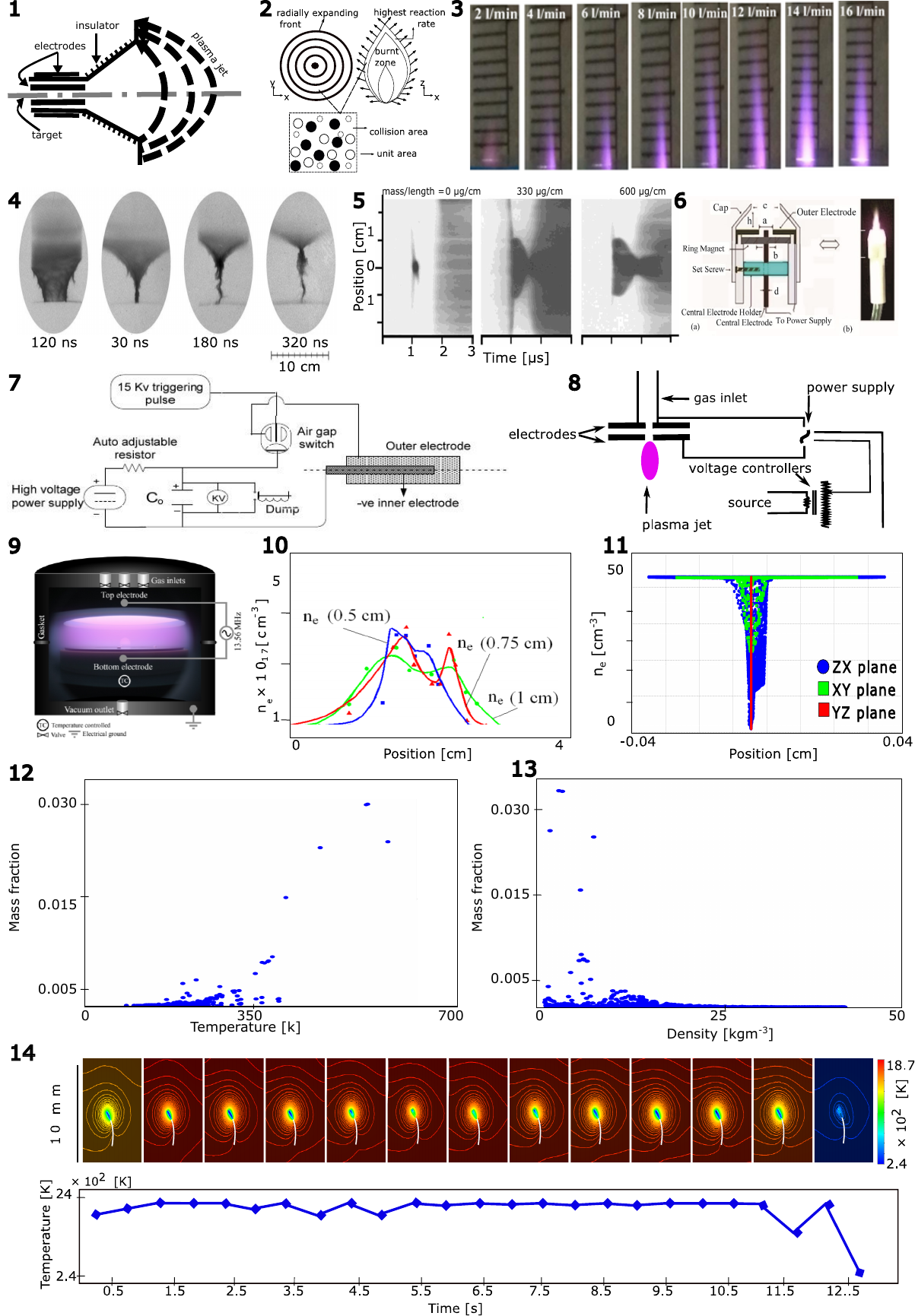}
        \caption{\textbf{1}, Principle of plasma jet generation.
        \textbf{2}, Principle of candle flame.
        \textbf{3}, Plasma length variation with different inlet flow rates. (adapted from \cite{lotfy2017cold}).
        \textbf{4}, Plasma discharges in pure neon (adapted from \cite{plasma4030033}).
        \textbf{5}, Effect of liners in PF discharge 
        (adapted from \cite{plasma4030033}).
        \textbf{6}, An air plasma generation setup (adapted from \cite{kuo2012air}).
        \textbf{7}, Circuit design of a 1.5 kJ coaxial plasma device (adapted from \cite{allam2011plasma}).
        \textbf{8}, Electric circuit of nitrogen-plasma jet device (adapted from \cite{lotfy2017cold}).
        \textbf{9},  Real-time plasma from an RIE setup (adapted from \cite{ghosh2016nanoscale}
        \textbf{10}, Electron density profiles at 3 different distances from the anode plane (adapted from  \cite{plasma4030033}).
        \textbf{11}, FEM simulated electron density vs position of candle flame.
        \textbf{12}, Temperature dependent oxygen mass fraction simulation in candle flame.
        \textbf{13}, Oxygen mass fraction vs density in candle plasma.
         \textbf{14}, Time-dependant temperature distribution over area of a candle plasma.}
        \label{Figure:2}
    \end{figure}

Plasma science is another sophisticated field coined by Langmuir\cite{langmuir1928oscillations} but the implementation towards realistic application seems far-fetched problem, unlike nanomaterials. 
A pair of electrodes and gases with electrical control systems is the basic requirement to create a controlled plasma.  
Its shape gives the mean electron density to quantify (or \textit{sense}) the concentration of gases, such as oxygen (Figure \ref{Figure:2}.{1}).
The diffusion and collision of molecular species in a plasma system contribute to its shape and orientation as show in a candle flame, Figure \ref{Figure:2}.{2}.
Generation of plasma variants is dependent on density, scattering energy, and temperature.
Changes in electrode geometry influence plasma generation significantly.
Oxygen concentration in a closed system is strongly dependent on temperature and density of plasma.
The Thomson cross-section for scattering of light by electrons ($\sigma_T = 0.665 \times 10^{-29}$ m$^2$) quantifies the radiation emitted from an electron in different directions to identify elemental species.
Thus, electron density trajectories (e.g., the solar corona with polynomial expressions) should be a parameter for oxygen sensing in PoC devices.
In a candle flame, the intense combustion reactions attributes to its highest temperature region (2200 K)\cite{zheng2019measurement} and should provide the amount of available oxygen. 

A large number of consistent plasma discharges (like 6.5 $\times$ 10$^6$) can be produced within a small energy range of 0.006 J to 0.10 J using a `coaxial and parallel-rail' electrode geometry\cite{guman1968solid}. 
Plasma focusing devices (PF) operating in the energy range of 10$^6$ J to $10^7$ J\cite{soto2005new} should be utilised towards the miniaturisation of oxygen sensors. 
Coupling electrons with plasma instabilities by laser contribute to the effect of magnetic field in plasma temperature and density \cite{nuckolls1972laser}.
Figure \ref{Figure:2}.{3} shows the variation of cold plasma with the inlet flow of $N_2$ gas in plasma jet device\cite{lotfy2017cold}.
It is functional at room temperature and produce reactive oxygen and nitrogen species, thus, suitable for wearable applications.
At a constant input voltage of 3 kV, a plasma jet length increased initially with the nitrogen flow rate ranging from 2 l/min to 16 l/min until it reached a steady state of 7 mm.
Figure \ref{Figure:2}.{4} gives the frame camera pictures from pure neon PF discharge. 
The purity of air can be found from the emission spectrum of a working gas discharge in air plasma\cite{5175412}. 
Another constraint in plasma discharges is liners as shown in Figure \ref{Figure:2}.{5}.
These highly dense plasma from PF devices act as a driver for the magnetic compression of liners which can be used for oxygen plasma as well\cite{fortov2002study}.
In a non-equilibrium air plasma produced for medical applications, the emission spectrum corresponding to radiation of 777.4 nm at the plasma plume showed that the plasma flux carried a high concentration of meta-stable atomic oxygen, which extended up to 30 mm from the source (cap) -- Figure \ref{Figure:2}.{6}\cite{kuo2012air}.
Improvised Langmuir probe method with specific electrode geometry gives a 2.5 times higher plasma density (where cathodic electrons ionise the metal to produce electrons and metal ions) and propagation speed\cite{tian2019discharge}.
The plasma arc diameter varies with different electrode material.
plasma arc produced between copper electrode and copper work piece at 12 A and 50 $\mu$m discharge gap width is found to be 251 $\mu$m.
Similarly for copper-zinc, copper-steel, steel-steel, tungsten-steel, it was found to be 300 $\mu$m, 233 $\mu$m, 270 $\mu$m, and 300 $\mu$m, respectively\cite{Li2020}. 
The plasma arc with 0.5 mm diameter was measured between copper electrodes with discharge gap width 0.1 mm  and discharge current of 17 A for 80 $\mu$s time duration\cite{kojima2008spectroscopic}.

\subsection{Towards miniaturisation} 
To construct and design portable plasma systems for PoC, we should investigate large-scale plasma systems.
The PF devices are characterised by gas breakdown, formation of current sheath and its rapid collapse by the action of Lorentz force  within a period of few microseconds\cite{scholz2000pf}. 
Applying this concept\cite{allam2011plasma}, in Figure(\ref{Figure:2}.{7}) studied the orientation and direction of plasma current sheath in a 1.5 kJ coaxial plasma device by the effect of $N_2$ gas within the pressure range of 1 Torr to 2.2 Torr. 
Lotfy et al. designed and constructed a nitrogen jet plasma device Figure (\ref{Figure:2}.{8}) which could produce cold plasma extending up-to 7 mm from the separation of electrodes\cite{lotfy2017cold}.
Figure \ref{Figure:2}.{9} depicts a real time plasma system under vacuum \cite{ghosh2016nanoscale}.
In such an electrically controlled plasma system, the electron density and other fundamental parameters share similar responses.
If we go back to PF devices, the electron density distribution with respect to position in a plasma sheath as in Figure  \ref{Figure:2}.{10} has a strong resemblance with the electron density distribution of a candle flame system as shown in Figure \ref{Figure:2}.{11}.
The density slowly rises at the initial stage because of the atomic photoionisation, and reaches a maximum peak that is around $3 \times 10^{17}$ cm$^{-3}$ for distances 0.5 cm, 0.75 cm, and 1 cm, from anode plane and abruptly collapses later. 
We obtained this distribution from a transient simulation of pressure-based turbulence model of a candle flame using finite element analysis solver (ANSYS Fluent 2021 R1 \cite{stolarski2018engineering}). 
This shows a miniatured versions of plasma systems for highly sensitive to molecular parameters. 

\subsection{Parameters of plasma systems for oxygen sensing}
Mass fraction of oxygen sensing dependent parameters are temperature, electron density, and region of highest reaction rates. 
We studied a flat braid wick candle in an adiabatic enclosure. 
The time-dependent simulation was carried out for determining the variations in oxygen concentration with time.
Figure \ref{Figure:2}.{12} and Figure \ref{Figure:2}.{13} shows the dependence of mass fraction of oxygen on temperature and density, respectively in our FEM simulation, which has significant similarities with \cite{cooper1966plasma}) temperature dependence on the amount of dielectric concentration in plasma.
The concentration of oxygen is maximum (at around 650 K temperature) due to incomplete and slow reaction-collision processes involving oxygen. 
It declines after 650 K due to vigorous reactions at the flame border area in a flame system.
As the combustion products increase over time, it pushes away the existing oxygen.
The maximum oxygen mass fraction is above 0.030 at around 2 kg/m$^3$ and approaches a minimum steady value of 0.005 after 10 kg/ m$^3$.
Solving the governing rate equations gives the population densities and relaxation times necessary to obtain a steady state in hydrogen plasma\cite{drawin1969influence}).
Hence, density is inversely related to the mass fraction of oxygen in the simulated combustion system, which is illustrated in Figure \ref{Figure:2}.{13}.

\subsection{Plasma for oxygen sensing}
Let us make the path clear towards oxygen sensing from the existing wider perspective.
Usually in plasma systems, the radiation loss with time can be estimated from the coronal approximation where the thermal limit is close to the ionisation limit\cite{cooper1966plasma}).
Here, the radiative decay rates are higher in compared to the collision decay rates and the transition probability $A(P)$ from the excited level $(P)$ to ground level $(P=1)$ will be high. 
Thus, the excited state of electrons will be short lived and are responsible for the thin optical regime of plasma.
In electrical discharge machining, the initial step towards the formation of arc plasma is ionisation and breakdown of the dielectric medium, such as oxygen. 
It was found that the diameter of molten region decreases even though the plasma arc diameter increases. 
The degree of ionisation of a molecular specie depends on the plasma temperature (over 5000 K) from which the plasma arc area and the rate coefficients for various recombination-collision reactions are estimated\cite{kojima2008spectroscopic}. 
In a plasma system, the molten area is categorised as region above 1808 K and the heat affected area is above 600 K based on the physical changes of electrodes, which confirms the effect of electrode material in plasma diameter\cite{Li2020}.
It also shows variation of dielectric medium, such as oxygen and its composition alters the plasma arc.
The interaction of electrons with electromagnetic waves can vary the refractive index of plasma and determine its electron density\cite{equipe1978tokamak}.

In our simulated model of a candle flame, the highest temperature is around 2400 K, which makes the physics of molten region very significant for oxygen sensing.
Here, various species collide with air molecules to form the emission spectra\cite{yambe2016investigation}. 
Thus, comparing and evaluating the molten region before and after the dielectric discharge with respect to the line pair method\cite{griem2005principles}) can be useful for measuring a particular constituent in the system.
In a multi-step combustion mechanism with known operating conditions (such as H$_2$, O$_2$, HO$_2$, H$^+$, OH$^-$, HO$_2$, O$^{2+}$, and $\delta$ is the third body in our simulation), the concentration of oxygen is found from the rate equations of other species involved in the system\cite{griffiths2019flame}.
The candle flame simulation in Figure \ref{Figure:2}.{14} gives the rate of product emissions and rate of oxygen with time dependency on the species considered for combustion.
In more complex burning mechanisms, the mass fraction of various components in the system at a given time can be estimated from the added species and possible products formed.
The density of excited states are less than the free electrons and bare nuclei densities because, the relaxation time of excited levels are comparatively small \cite{bates1962recombination}, the energy is generated as radiation when an electron recombines with ionised atom and this process can be enhanced in the presence of oxygen.
In Figure \ref{Figure:2}.{14}, the emissivity is lower near the wick and increases radially towards the flame border.
The flame border is the region of highest reaction rates and the emissivity has an inverse relation to wavelength\cite{zheng2019measurement}. 
Plasma-based oxygen sensors can be regarded as an efficient sensing methodology due to its modus-operandi and accessibility.

\section{Discussion}
We found that developments of oxygen sensing methods from various states of matter are studied with optical, electrical, and chemical means. 
Oxygen can be determined at the molecular level with single-molecule level accuracy in real-time at fast timescale. 
Nanomaterial-based oxygen sensors give accurate measurements on concentration of molecular oxygen. 
Lack of safe disposal methods and non-unified policies of hazardous nanomaterials wastes have been a concern. 
Ranking nanomaterial according to their performance, TiS$_2$, SiBi, graphene, and zeolite will top the list.
Cost of nanomaterial production are decreasing with scalability but solid-state nanofabrications have been expensive, which needs significant attentions. 
Plasma-based oxygen sensing is a high-throughput and cost effective method. 
It eliminates major organic contaminants and produce less pollutants but high-energy emissions from plasma should be fine-tuned with energy filters for wearable applications.
Giant plasma systems and small confined low-temperature plasma systems share evident similarities in terms of electron density and temperature to the concentration ratio of specific molecular species, which can be adapted for developing oxygen sensors. 
Candle-based oxygen systems have low-energy emissions, and cold-plasma systems at room-temperature are suitable for wearables.
Another cost effective sensing methodology is solid-state-material-based nanofluidics, which will optimally combine nanomaterials and plasma together to achieve a fast real-time oxygen sensor, which can be analysed with cutting-edge in-situ super-resolution microscopy, electron energy loss spectroscopy, and x-ray diffraction.  
Integrating oxygen concentrator with artificial neural network, lab-on-chip, smartphone technology, and unconventional computing is a way forward to monitor patients with unmanageable health issues, like COVID-19, hypoxia, and resistant pneumonia. 
Our perspective will give the insight to the startups, unicorns, Governments, philanthropists, and financial institutions to embark upon creating low-cost oxygen sensing PoCs for creating a strong health-care friendly sustainable economy.

\subsection*{Permission to Reuse and Copyright}
For adaptation of images, authors obtained permission from the  authors of the corresponding papers.

\section*{Conflict of Interest Statement}

Author MG is employed by the Thermo Fisher Scientific. Authors VJ and KVC are employed by the International Centre for Nanodevices. The remaining authors declare that the research was conducted in the absence of any commercial or financial relationships that could be construed as a potential conflict of interest. 

\section*{Author Contributions}
VJ and KVC collected all the data, analysed them, and written the manuscript.
MNCV and VK performed an extended experiment, which supported writing this manuscript. 
MG, GMR, and SG supervised the research and partly written the manuscript.
SG proposed the research.
All other have read and agreed upon the scientific views of the perspective.

\section*{Funding}
Details of all funding sources should be provided, including grant numbers if applicable. Please ensure to add all necessary funding information, as after publication this is no longer possible.

\section*{Acknowledgements}
Authours are thankful to Jintu James, Midhun George Thomas, and Subham Ghosh for productive discussion. The study was supported by internal funding of the Open Academic Research Council and the International Centre for Nanodevices. All the authors are extremely grateful to Sagar Gosalia for managing the research facilities.

\bibliography{Main.bib}

\end{document}